\documentclass[aps,twocolumn,prc,superscriptaddress]{revtex4}

\usepackage{graphicx}  % Include figure files
\usepackage{subfigure}
\usepackage{multirow}

\usepackage{fancyhdr}
\usepackage{longtable}
\usepackage{parskip}
\usepackage[T1]{fontenc}
\usepackage{dcolumn}   % Align table columns on decimal point
\usepackage{bm}        % bold math
\usepackage{amsfonts}  % Common math fonts
\usepackage{amsmath}   % Common math functions
\usepackage{amssymb}   % Common math symbols
\usepackage{url}
\usepackage{color}
\usepackage{ulem}
\usepackage{hyperref}
\hypersetup{
    colorlinks=true,
    linkcolor=blue,
    filecolor=blue,
    urlcolor=blue,
    citecolor=blue,
}
\newcommand{\beq}{\begin{equation}}
\newcommand{\eeq}{\end{equation}}
\newcommand{\beqa}{\begin{eqnarray}}
\newcommand{\eeqa}{\end{eqnarray}}

\newcommand{\pwisein}{\left\{ \begin{array}{ll}}
\newcommand{\pwiseout}{\end{array}\right.}
\newcommand{\ket}[1]{\left| #1 \right\rangle}
\newcommand{\braket}[2]{\left\langle #1 | #2 \right\rangle}
\newcommand{\mel}[3]{\left\langle #1\left|#2 \right| #3 \right\rangle}
\newcommand{\abs}[1]{\left| #1 \right|}

\begin{document}

\title{Quantum theory of isomeric excitation of $^{229}$Th in strong laser fields}

\author{Wu Wang}

\affiliation {\it Beijing Computational Science Research Center, Beijing 100193, China}

\author{Xu Wang}
\email{xwang@gscaep.ac.cn}
\affiliation {\it Graduate School, China Academy of Engineering Physics, Beijing 100193, China}

\date{\today}

\begin{abstract}

A general quantum mechanical theory is developed for the isomeric excitation of $^{229}$Th in strong femtosecond laser pulses. The theory describes the tripartite interaction between the nucleus, the atomic electrons, and the laser field. The nucleus can be excited both by the laser field and by laser-driven electronic transitions. Numerical results show that strong femtosecond laser pulses are very efficient in exciting the $^{229}$Th nucleus, yielding nuclear excitation probabilities on the order of $10^{-11}$ per nucleus per pulse. Laser-driven electronic excitations are found to be more efficient than direct optical excitations.

\end{abstract}

\maketitle

\section{Introduction}

Nuclear isomers are nuclear metastable states with relatively long half-lives. They have important applications in energy storage \cite{Walker_1999_Nature, Carroll_2001_Xray_relaaseenergy, Zadernovsky2002, Palffy_2007_PRL_TriggerNEEC, Yuanbin_PRL_2019}, medical imaging \cite{Cutler2013}, nuclear structure elucidation \cite{Walker_2020}, etc. One of the most fascinating nuclear isomers of substantial recent interest is the $^{229}$Th isomer, which has an extremely low energy of only about 8 eV above the nuclear ground state \cite{K_R_NPA_1976, Reich-90, Helmer_PRC_1994, Beck_PRL_2007, Seiferle2019, Wense2016, Thielking2018, Minkov_2019_PRL, Yamaguchi_2019_PRL, Sikorsky_2020_PRL}. It has been proposed as a ``nuclear clock'' \cite{Peik_2003, Peik-09, Rellergert-10, Campbell_2012_PRL_clock, Flambaum-06, Berengut-09, Fadeev-20} that may outperform or complement today's atomic clocks.

One of the current research focuses is to find efficient methods to excite the $^{229}$Th nucleus from the ground state to the isomeric state \cite{Jeet_PRL_2015, Yamaguchi_NewJPhys_2015, Peik-15, Stellmer_PRA_2018, Masuda_2019_nature_xray, Tkalya-00, Tkalya-20, Zhang-22, Zhang-23, Tkalya-92, Porsev_2010_PRL_TEB, Borisyuk_2019_PRC_EBcontinuum, Bilous_2020_PRL_Ions, Nickerson_2020_PRL_Nuclear, Bilous_2018_PRC, Feng-22, Qi-23}. Existing methods or proposals may be summarized into the following categories: optical excitation (OE), electronic excitation (EE), or laser-driven electronic excitation (LDEE). OE using vacuum ultraviolet light around 8 eV is conceptually straightforward, however, several experimental attempts have given negative results \cite{Jeet_PRL_2015, Yamaguchi_NewJPhys_2015, Peik-15, Stellmer_PRA_2018} possibly due to inaccurate knowledge of the isomeric energy. An indirect OE method was demonstrated \cite{Masuda_2019_nature_xray} using 29-keV synchrotron radiations to pump the nucleus to the second excited state which then decays preferably into the isomeric state \cite{Tkalya-00}. Several EE processes are discussed and calculated, including nuclear excitation by inelastic electron scattering \cite{Tkalya-20, Zhang-22}, by bound electronic transition or by electron capture \cite{Zhang-23}. For the LDEE category, electronic bridge (EB) schemes are most discussed \cite{Tkalya-92, Porsev_2010_PRL_TEB, Borisyuk_2019_PRC_EBcontinuum, Bilous_2020_PRL_Ions, Nickerson_2020_PRL_Nuclear}. The EB method has not been experimentally realized due to the requirements on resonant conditions for both nuclear and electronic transitions.

The possibility of using strong femtosecond laser pulses for the isomeric excitation has been considered by us previously \cite{Wu_PRL_2021, Wang-22}. We explain that efficient isomeric excitation of $^{229}$Th can be achieved through a laser-driven electron recollision process: (i) An outer electron is pulled out by the strong laser field; (ii) The electron is driven away but it has a probability to be driven back and ``recollide'' with its parent ion core when the oscillating laser field reverses its direction; (iii) The recolliding electron excites the $^{229}$Th nucleus from the ground state to the isomeric state. This electron recollision process is well known \cite{Kulander_Plenum_1993, Schafer-93, Corkum-93} and it is the core process underlying strong-field phenomena including high harmonic generation \cite{McPherson_1987, Ferray_1988, Seres2005}, nonsequential double ionization \cite{Walker-94, Palaniyappan-05, Becker-12}, laser-induced electron diffraction \cite{Morishita-PRL-2008,Blaga-Nature-2012,Wolter-Science-2016}, attosecond pulse generation \cite{Krausz-09, Zhao-12, Li-17, Gaumnitz-17}, etc. Therefore this recollision-induced nuclear excitation (RINE) process is an interesting combination of $^{229}$Th nuclear physics and strong-field atomic physics \cite{Andreev_PRA_2019,Wu_JPB_2021}.
Semiclassical calculations show that the probability of isomeric excitation is on the order of 10$^{-12}$ per femtosecond laser pulse per nucleus \cite{Wang-22}.

In this paper we go further by developing a quantum mechanical theory for the isomeric excitation process by strong femtosecond laser fields. Such a theory is desirable for a couple of reasons. First, it provides a quantum basis for the RINE process and benchmarks for our previous semiclassical calculations. Second, it provides a more complete picture of isomeric excitation in strong laser fields by including processes beyond RINE. The RINE involves only laser-driven free-free electronic transitions, whereas the quantum theory also includes intrinsically laser-driven free-bound and bound-bound electronic transitions. The quantum theory describes the tripartite interaction between the nucleus, the atomic electrons, and the laser field. It encloses simultaneously OE and LDEE channels, so direct comparisons between these channels are possible for given laser parameters.

This paper is organized as follows. In Sec. II the tripartite quantum theory is developed. Numerical results of nuclear excitation probabilities under different laser parameters are given in Sec. III. Further discussions are given in Sec. IV, and a conclusion is given in Sec. V.

\section{The Quantum Theory}

\subsection{The tripartite interaction}

\begin{figure}[t!]
    \centerline {\includegraphics[width=5.6cm]{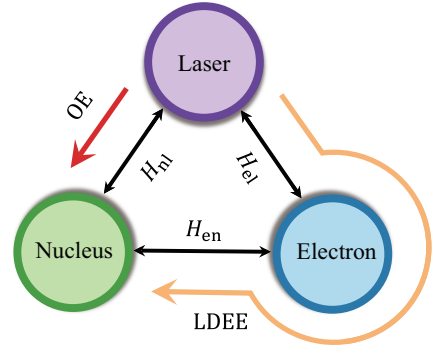}}
    \caption{Illustration of the tripartite interaction between the nucleus, the atomic electrons, and the laser field.}
    \label{f_model}
\end{figure}

The Hamiltonian of the laser-nucleus-electron system can be written as
\begin{equation}\label{2022_5_21_1}
  H = H_{\mathrm{e}}+H_{\mathrm{n}}+H_{\mathrm{en}}+H_{\mathrm{el}}+H_{\mathrm{nl}}~,
\end{equation}
where the Hamiltonians on the right hand side are, respectively, for the atomic electrons, the nucleus, the electron-nucleus coupling, the electron-laser coupling, and the nucleus-laser coupling. They describe the tripartite interaction between the nucleus, the atomic electrons, and the laser field, as illustrated in Fig. \ref{f_model}.
The state space of $H_{\mathrm{e}}$ is the atomic eigenstates, including both bound and continuum states, which are denoted by $\ket{\varphi_i}$. The state space of $H_{\mathrm{n}}$ contains the nuclear ground state $\ket{I_g M_g}$ and the isomeric excited state $\ket{I_e M_e}$. The electron-nucleus coupling $H_{\mathrm{en}}$ is given as a summation over irreducible tensor operators \cite{Schwartz_1957}
\begin{equation}\label{2022_5_21_2}
  H_{\mathrm{en}}=\sum_{\tau={E},{M}}\sum_{lm}(-1)^m\mathcal{M}^{\tau}_{l-m} T^{\tau}_{lm} ~,
\end{equation}
where $\mathcal{M}^{\tau}_{lm}$ is the nuclear multipole moment of type $\tau$ ($E$ for electric, $M$ for magnetic) and rank $l$. $T^{\tau}_{lm}$ is the corresponding multipole operator for the atomic electrons and is given by
\begin{equation}\label{2022_3_29_1}
  \begin{split}
     T^{E}_{lm} =& \sqrt{\frac{4\pi}{2l+1}}\int\frac{\rho_\mathrm{e}(\bm r) }{r^{l+1}}Y_{lm}(\theta,\phi)\mathrm{d}\tau,\\
     T^{M}_{lm} =  &\sqrt{\frac{4\pi}{2l+1}}\int\frac{i}{cl}\frac{\bm j_\mathrm{e}(\bm r)\cdot \bm L[Y_{lm}(\theta,\phi)]}{r^{l+1}}\mathrm{d}\tau,
  \end{split}
\end{equation}
where $Y_{lm}(\theta,\phi)$ is spherical harmonics, $\rho_\mathrm{e}(\bm r)$ and $\bm j_\mathrm{e}(\bm r)$ are charge and current density operators for the electron.

The nucleus-laser coupling is given by
\begin{equation}\label{2022_3_29_2}
   H_{\mathrm{nl}}=-\frac{1}{c}\int \bm j_n(\bm r)\cdot\bm A(\bm r,t)\mathrm{d}\tau,
\end{equation}
where $\bm j_n(\bm r)$ and $\bm A(\bm r,t)$ are the operators for the nuclear current density and the laser vector potential, respectively. The vector potential $\bm A(\bm r,t)$ satisfies the Coulomb gauge $\bm\nabla\cdot \bm A(\bm r,t)=0$. Assume the vector potential has the following form
\beq
\bm A(\bm r,t)=\frac{\hat{\bm z}}{2} \big[ w(t)\mathrm{exp}(i\bm k\cdot\bm r) + \mathrm{c.c.}\big],
\eeq
where $w(t)$ is a temporal function, $\bm k$ and $\hat{\bm z}$ are the wave vector and the polarization unit vector. By expanding the vector potential in vector spherical harmonics, the nucleus-laser coupling can be rewritten as
\begin{equation}\label{2022_5_24_1}
  H_{\mathrm{nl}}=-\sqrt{4\pi}\sum_{\tau={E},{M}}\sum_{lm}\sqrt{2l+1}(-1)^m\mathcal{M}^{\tau}_{l-m}C_{lm}^\tau~,
\end{equation}
where $C_{lm}^\tau$ is the multipole expansion coefficient of the vector potential
\begin{equation} \label{e.Ctaulm}
   \begin{split}
  C_{lm}^E     & =\frac{ik^{l-1}k_0}{(2l+1)!!}\sqrt{\frac{l+1}{l}} \big[w(t)i^{l-1}+\mathrm{c.c.}\big]\frac{\hat{\bm z}}{2}\cdot\bm A_{lm}^\mathrm{E}(\hat{\bm k})~,  \\
   C_{lm}^M    &=  \frac{-ik^{l}}{(2l+1)!!}\sqrt{\frac{l+1}{l}}\big[w(t)i^{l}+\mathrm{c.c.}\big]\frac{\hat{\bm z}}{2}\cdot\bm A_{lm}^\mathrm{M}(\hat{\bm k})~.
   \end{split}
\end{equation}
Here $k_0=\omega_0/c$ with $\omega_0$ the nuclear energy gap, $\hat{\bm k}=\bm k/ k$ is the unit vector along the $\bm k$ direction. $\bm A_{lm}^\mathrm{\tau}(\hat{\bm k})$ is the transverse vector spherical harmonics \cite{W.J_Aphy_2007}
\begin{equation}\label{2022_3_29_3}
  \begin{split}
     \bm A_{lm}^E(\hat{\bm k}) = &    \frac{k}{\sqrt{l(l+1)}}\bm \nabla Y_{lm}(\hat{k}),   \\
   \bm A_{lm}^M(\hat{\bm k}) =   & \frac{1}{\sqrt{l(l+1)}}\bm LY_{lm}(\hat{k}),
  \end{split}
\end{equation}
with the orthonormal relation
\begin{equation}\label{2022_3_29_4}
  \int \bm A_{lm}^\mathrm{\tau}(\hat{\bm k})\cdot\bm A_{l^{\prime}m^{\prime}}^\mathrm{\tau^{\prime} *}(\hat{\bm k})\mathrm{d}\Omega_{\hat{\bm k}}=
  \delta_{ll^{\prime}}\delta_{mm^{\prime}}\delta_{\tau \tau^{\prime}}.
\end{equation}

Similar to the nucleus-laser coupling, the electron-laser coupling can also be expressed as a summation of multipole terms. However, the dipole approximation is usually sufficient, in which only the electric dipole interaction is used to describe the electron dynamics in the laser field. In this approximation, the electron-laser coupling is written as
\begin{equation}\label{2022_3_29_5}
    H_{\mathrm{el}}=- \bm D \cdot \bm E(t).
\end{equation}
Here $\bm D$ is the dipole moment operator, and $\bm E(t)$ is the laser electric field at the position of the atom: $\bm E(t) \equiv \bm E(\bm r = \bm 0, t) = - \partial \bm A(\bm r = \bm 0, t) / \partial t$.

\subsection{The nuclear excitation probability}

Without the laser field, the Hamiltonian of the nucleus-electron system is
\begin{equation}\label{2023_4_24_1}
  H_0 = H_{\mathrm{e}}+H_{\mathrm{n}}+H_{\mathrm{en}},
\end{equation}
with eigenstates
\begin{equation}\label{2022_3_29_6}
  H_0 \ket{\Psi_{\mu,\varepsilon}}=E_{\mu,\varepsilon}\ket{\Psi_{\mu,\varepsilon}},
\end{equation}
where $\mu$ ($\varepsilon$) denotes the nuclear state (electronic state). For example, $\mu = g$ or $e$ for the nuclear ground state or the isomeric excited state, and $\varepsilon = i$ or $f$ for the initial or the final electronic state. The eigenstate $\ket{\Psi_{\mu,\varepsilon}}$ can be expanded using the uncoupled states $\ket{I_\mu M_\mu,\varphi_\varepsilon}\equiv \ket{I_\mu M_\mu}\otimes\ket{\varphi_\varepsilon}$ using perturbation theory
\begin{equation}\label{2022_3_29_9}
  \begin{split}
     \ket{\Psi_{\mu,\varepsilon}}= & \ket{I_\mu M_\mu,\varphi_\varepsilon}+\sum_{\mu^\prime,\varepsilon^\prime} \ket{I_{\mu^\prime } M_{\mu^\prime},\varphi_{\varepsilon^\prime}}
\times\\
 & \frac{\mel{I_\mu M_\mu,\varphi_\varepsilon}{H_{\mathrm{en}}}{I_{\mu^\prime } M_{\mu^\prime},\varphi_{\varepsilon^\prime}}}{E_\varepsilon-E_{\varepsilon^\prime}+E_\mu-E_{\mu^\prime}},
  \end{split}
\end{equation}
where $E_\varepsilon$ and $E_\mu$ are the energy eigenvalues corresponding to the atomic state $\ket{\varphi_\varepsilon}$ and the nuclear state $\ket{I_\mu M_\mu}$, respectively.

Initially at time $t_0$ the nucleus-electron system is assumed to be in its ground state $\ket{\Psi_{g,i}}$. Then the probability of nuclear isomeric excitation at a later time $t>t_0$ is given by
\begin{equation}\label{2022_5_22_1}
  P_{\mathrm{exc}}(t) = \sum_{f}|\mel{\Psi_{e,f}}{U(t,t_0)}{\Psi_{g,i}}|^2~,
\end{equation}
where $U(t,t_0)$ is the time evolution operator corresponding to the total Hamiltonian $H$, and the summation runs over all final electronic states.

The time evolution is computationally very demanding due to the large state space (which comes mostly from the electronic states). However, one notices that the couplings of the nucleus to the laser field and to the electrons are weak. This allows the usage of perturbative treatments for the laser-nucleus and electron-nucleus couplings, reducing the computation load substantially. The laser-electron coupling, in contrast, is very strong and must be treated non-perturbatively. Using the time-dependent perturbation theory, the time evolution operator $U(t, t_0)$ becomes
\begin{equation}\label{2022_3_29_7}
  U(t,t_0)=U_0(t,t_0)e^{-iH_{\mathrm{n}}(t-t_0)}[1-iV_I(t)],
\end{equation}
where $V_I(t)$ is defined by
\begin{equation}\label{2022_3_29_8}
\begin{split}
  V_I(t)=     \int_{t_0}^t & e^{-iH_{\mathrm{n}}(t_0-t')}U_0(t_0,t')(H_{\mathrm{en}}+H_{\mathrm{nl}}) \\
    & \times e^{-iH_{\mathrm{n}}(t'-t_0)}U_0(t',t_0)\mathrm{d}t'.
\end{split}
\end{equation}
In the above expressions, $U_0(t,t_0)$ is the time evolution operator corresponding to $(H_{\mathrm{e}}+H_{\mathrm{el}})$.

Substituting Eqs. \eqref{2022_3_29_9} and \eqref{2022_3_29_7} into Eq. \eqref{2022_5_22_1},
$P_{\mathrm{exc}}(t)$ can be derived into the following form after averaging over initial nuclear states and summing over final states:
\begin{equation}\label{2022_5_24_3}
   P_{\mathrm{exc}}(t)=4\pi\sum_{\tau,l}\frac{B(\tau l, g\rightarrow e)}{(2l+1)^2}\sum_{f,m}\abs{N^{\tau;fi}_{lm}}^2~.
\end{equation}
The favorable feature of the above formula is that the nuclear transitions are packed in $B(\tau l, g\rightarrow e)$ and the electronic transitions are packed in $N^{\tau;fi}_{lm}$. The former is the reduced nuclear transition probability
\begin{equation*}
   B(\tau l, g\rightarrow e)=\frac{2l+1}{4\pi(2I_g+1)}\sum_{M_eM_gm}\abs{\mel{I_e M_e}{\mathcal{M}^\tau_{lm}}{I_g M_g}}^2
\end{equation*}
and $N^{\tau;fi}_{lm}$, depending only on the electronic initial and final states, is given by
\begin{equation}\label{2022_5_23_1}
\begin{split}
   N^{\tau;fi}_{lm}  =  &- i\int_{t_0}^t\mel{\varphi_f(t')}{{T}_{lm}^\tau}{\varphi_i(t')}e^{i\omega_0 t'}\mathrm{d}t' \\
   &+ e^{i\omega_0t} \sum_{kn}\braket{\varphi_f(t)}{\varphi_n}\frac{\mel{\varphi_n}{{T}_{lm}^\tau}{\varphi_k}}{\varepsilon_n-\varepsilon_k +\omega_0}\braket{\varphi_k}{\varphi_i(t)} \\
   &+ e^{i\omega_0t_0}\frac{\mel{\varphi_f}{{T}_{lm}^\tau}{\varphi_i}}{\varepsilon_i-\varepsilon_f-\omega_0} \\
   &+ i\delta_{fi}\sqrt{4\pi(2l+1)} \int_{t_0}^tC_{lm}^\tau(t')e^{i\omega_0 t'}\mathrm{d}t'~.
\end{split}
\end{equation}
In the above formula $\ket{\varphi_{i/f}(t)}$ is the electronic state evolved by $U_0(t,t_0)$ from the state $\ket{\varphi_{i/f}}$. Note that both LDEE and OE channels emerge. The first three lines of Eq. (\ref{2022_5_23_1}) all contain the (time-dependent) electronic states and they describe the LDEE channel. As the atomic state space includes both bound and free states, the contributions from bound-bound, bound-free, and free-free electronic transitions to the nuclear excitation are taken into account intrinsically.
The last line, which does not contain electronic states, describes the OE channel, with the $C_{lm}^\tau$ given in Eq. (\ref{e.Ctaulm}). The OE channel does not change the electronic state, hence the $\delta_{fi}$.

For the $^{229}$Th nucleus, the leading nuclear transitions from the ground state to the isomeric state are magnetic dipole ($M1$) and electric quadrupole ($E2$). In our calculation we use the reduced transition probability values $B(M1, e\rightarrow g)=0.005$ W.u. and $B(E2, e\rightarrow g)=30$ W.u., as predicted recently by Minkov and P\'alffy \cite{Minko_PRC_2021}.

\subsection{The time-dependent ZORA equation}

The dynamics of the atomic electrons driven by a strong laser pulse has been extensively studied in strong-field atomic physics. Theoretically, most strong-field phenomena can be well understood by solving the time-dependent Schr\"odinger equation under a single-active-electron (SAE) approximation \cite{ADK, Kulander-87, Schafer-93, Corkum-93, Awasthi-08, Le-16}, which assumes that only the outermost electron actively responds to the external laser field, with the remaining electrons contributing a mean-field potential.

The difference between the current work and traditional strong-field atomic physics is the addition of the nuclear degree of freedom. We find that this difference {\it makes the Schr\"odinger equation insufficient}. The main contribution to nuclear excitation comes from electron wave functions very close (around 10$^{-2}$ a.u.) to the nucleus due to the factor $r^{-l-1}$ in the electronic operator $T_{lm}^\tau$ of Eq. \eqref{2022_3_29_1}. Yet the amplitudes of the Schr\"{o}dinger wave functions, even for low electron energies, can be very different from those of the Dirac wave functions in this region due to the high nuclear charge ($Z=90$). Therefore relativistic effects are important for the isomeric excitation. A similar conclusion has also been given in nuclear excitation by inelastic electron scattering \cite{Zhang-22}.

The straight way to calculate the nuclear excitation probability is to solve the time-dependent Dirac equation, which, however, is very time consuming. An alternative and much more economical approach is used in this paper. We adopt for $H_\mathrm{e}$ a so-called zero-order-regular-approximation (ZORA) Hamiltonian \cite{Chang_1986, Lenthe_1993, Lenthe_1994}, instead of the Dirac Hamiltonian. The ZORA Hamiltonian is an effective two-component Hamiltonian giving an accurate approximation of relativistic effects:
\begin{equation}\label{2022_5_25_1}
  H_\mathrm{ZORA}=\bm \sigma\cdot \bm p\frac{1}{2-\alpha^2V(r)}\bm \sigma\cdot\bm p +V(r),
\end{equation}
where $\bm \sigma$ is the Pauli matrix, $\alpha$ is the fine structure constant, and $V(r)$ is a central potential felt by the electron.

\begin{figure}[t!]
    \centerline {\includegraphics[width=5.6cm]{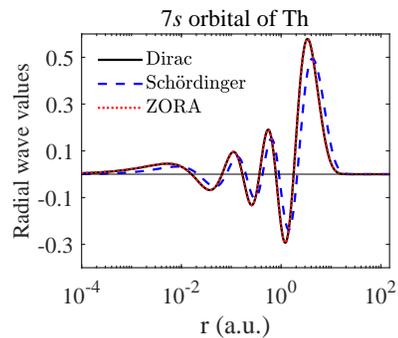}}
    \caption{Radial wave function of the 7$s$ orbital of the $^{229}$Th atom, calculated by three different Hamiltonians as labeled. The ZORA wave function is almost identical to the Dirac one, and the two can barely be visually distinguished.}
    \label{fig:2022_5_26_1}
\end{figure}

Figure \ref{fig:2022_5_26_1} shows the radial wave function of the 7$s$ orbital of the $^{229}$Th atom, for the Schr\"{o}dinger case, the (large component of the) Dirac case, and the ZORA case. One can see that the ZORA wave function is almost identical to the Dirac one, whereas the Schr\"{o}dinger wave function has smaller amplitudes close to the nucleus, leading to an underestimation of the nuclear excitation probability up to an order of magnitude. The potential energy $V(r)$ used in Fig. \ref{fig:2022_5_26_1} and in the results below is calculated by the RADIAL package \cite{radial_2019} based on a self-consistent Dirac-Hartree-Fock-Slater method.

\begin{figure*}[t!]
\centering
\subfigure{
\centering
\includegraphics[width=5.4cm]{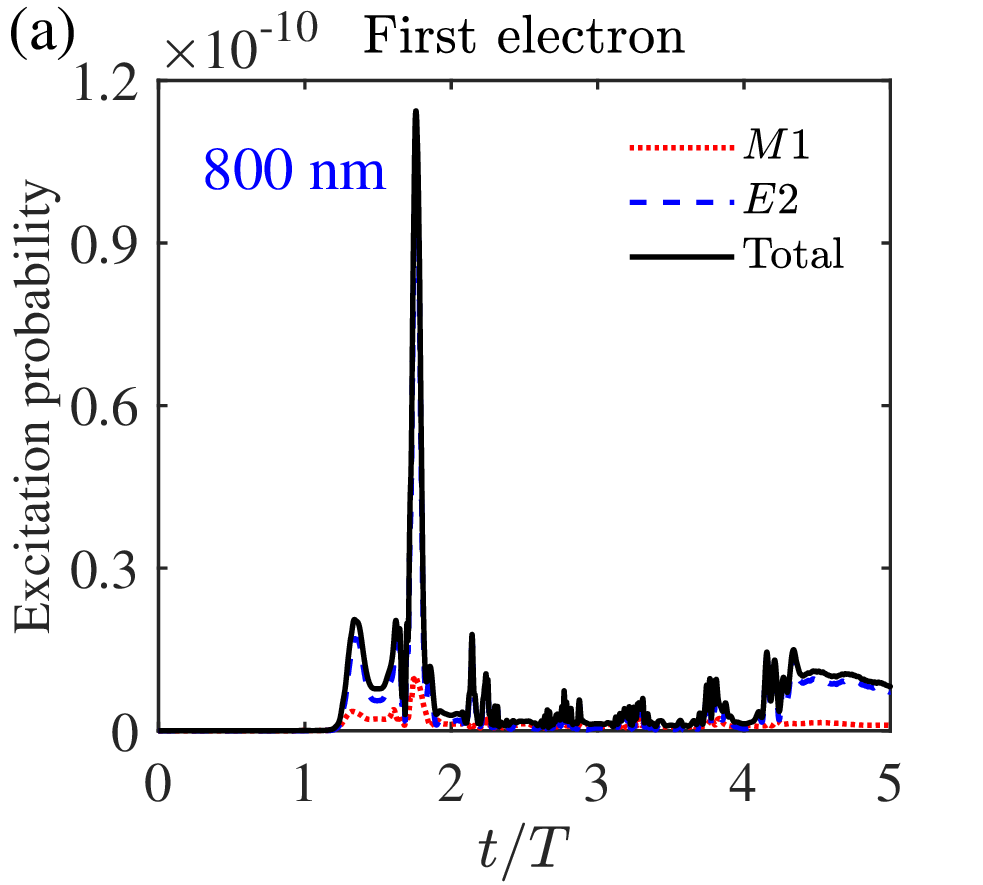}\label{Fig:four_2022_5_28_1_a}
%\caption{fig1}
}%
\subfigure{
\centering
\includegraphics[width=5.4cm]{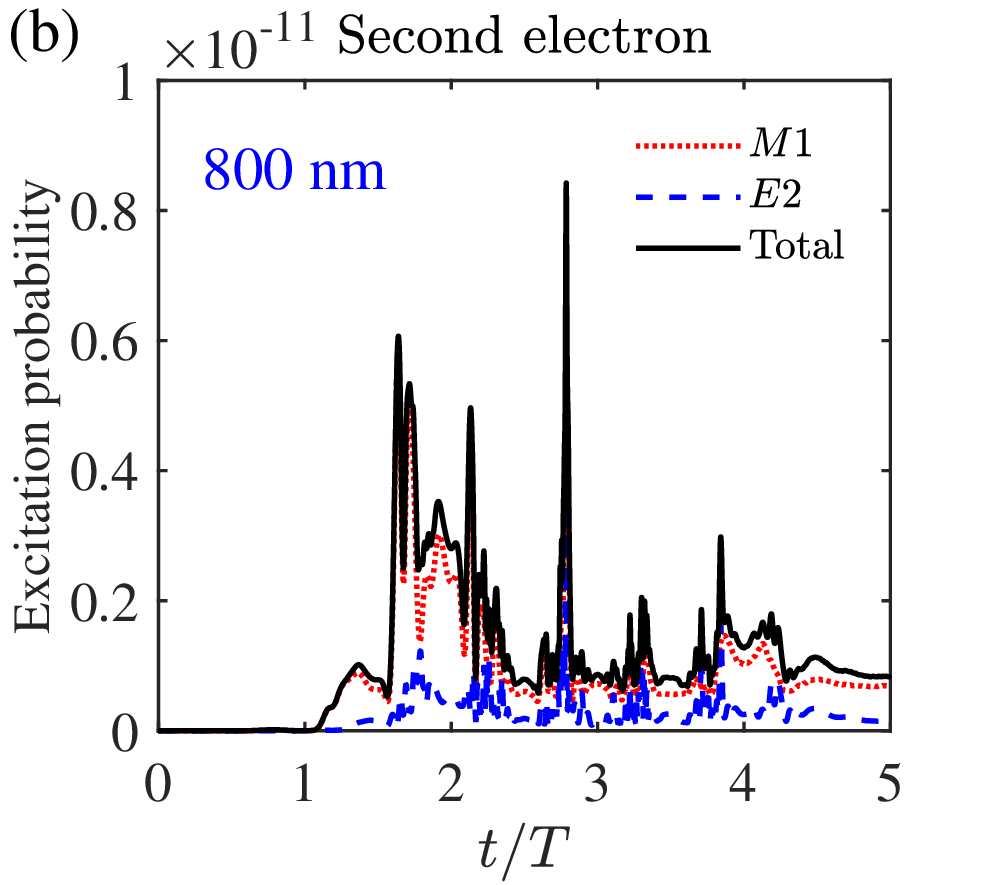}\label{Fig:four_2022_5_28_2_b}
%\caption{fig2}
}%
\subfigure{
\centering
\includegraphics[width=5.4cm]{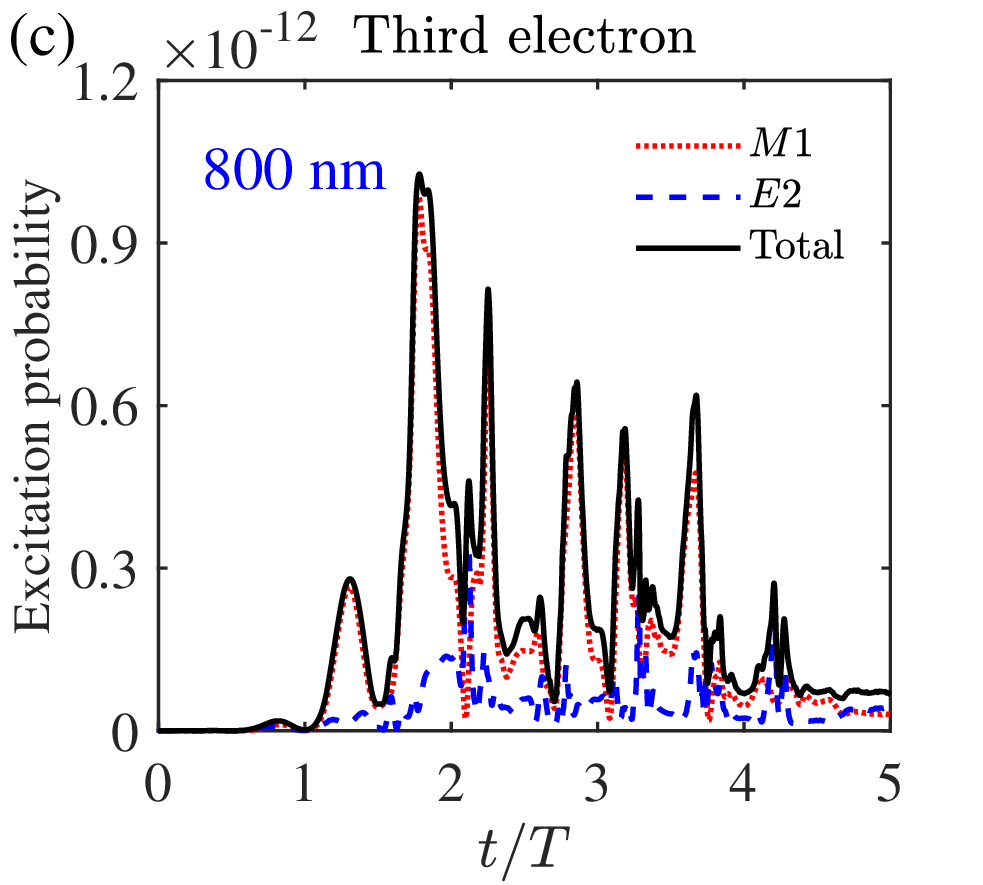}\label{Fig:four_2022_5_28_3_c}
%\caption{fig2}
}%

\subfigure{
\centering
\includegraphics[width=5.4cm]{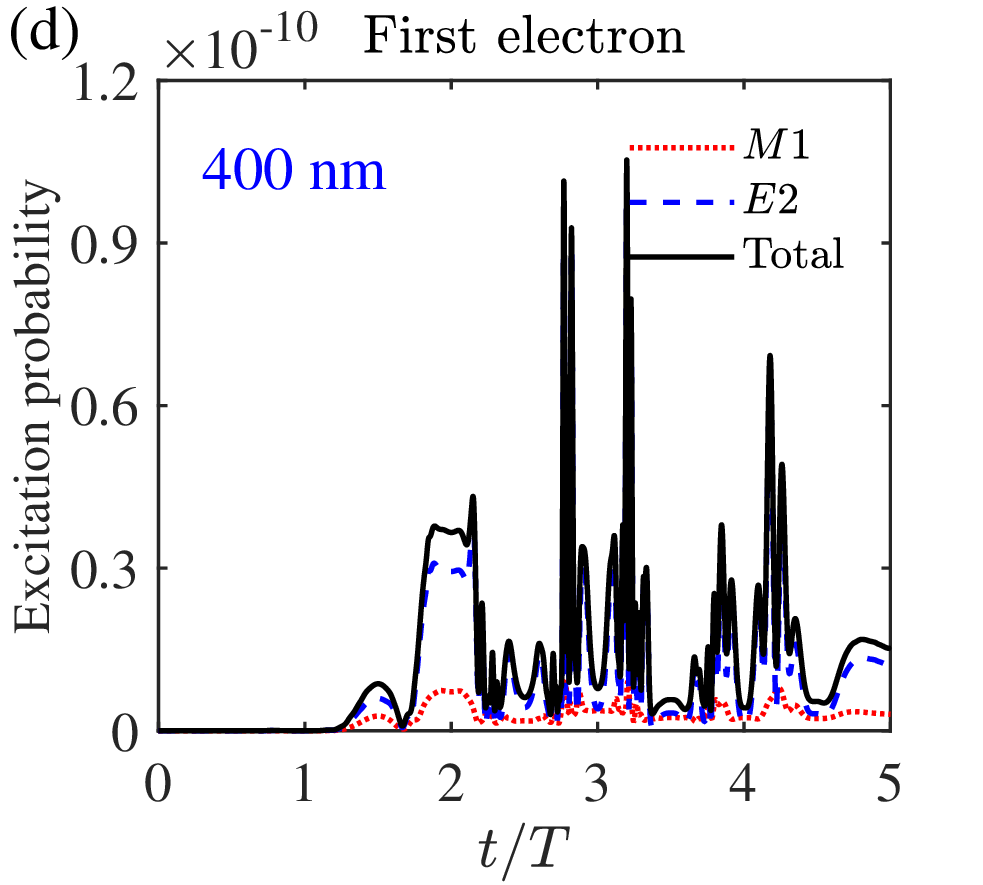}\label{Fig:four_2022_5_28_1_a}
%\caption{fig1}
}%
\subfigure{
\centering
\includegraphics[width=5.4cm]{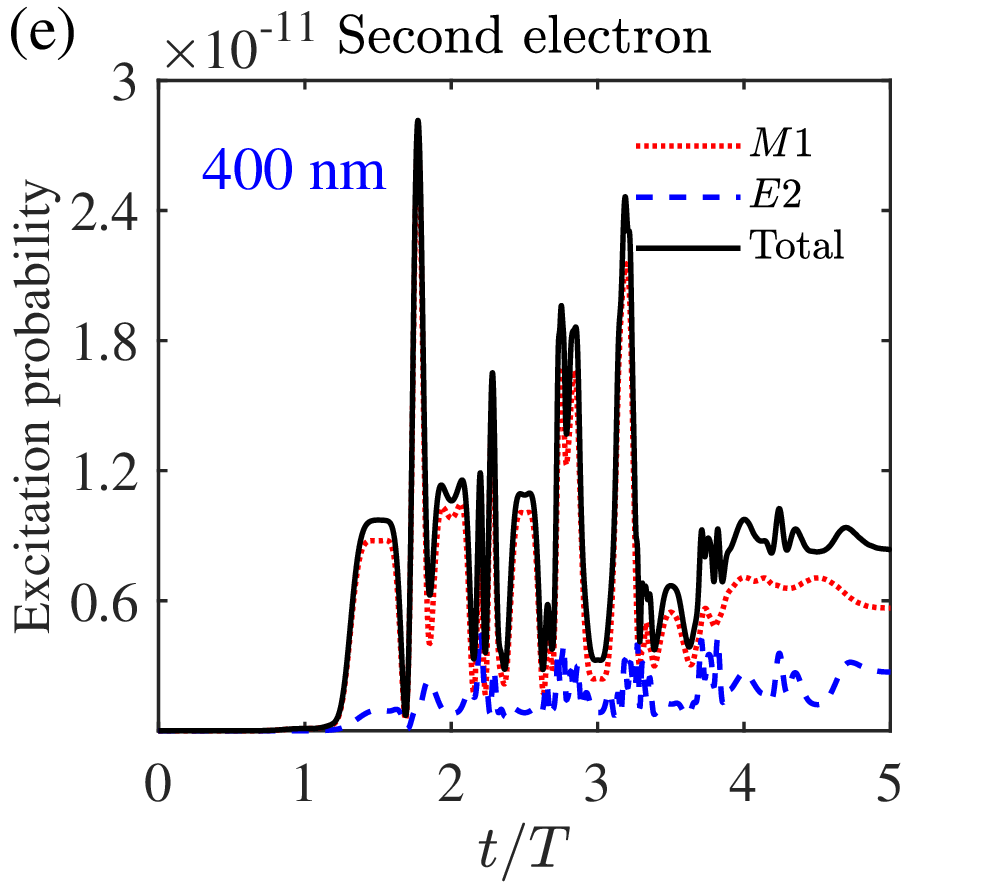}\label{Fig:four_2022_5_28_2_b}
%\caption{fig2}
}%
\subfigure{
\centering
\includegraphics[width=5.4cm]{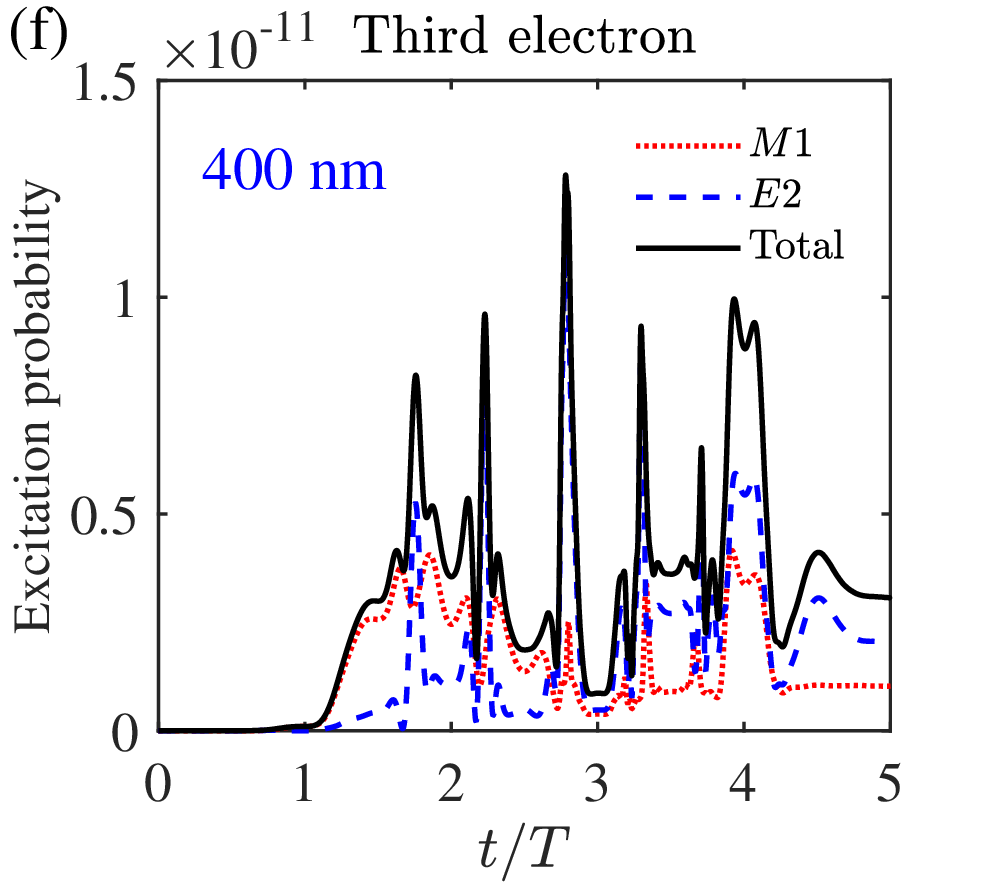}\label{Fig:four_2022_5_28_3_c}
%\caption{fig2}
}%

\centering
\caption{ \label{fig:2022_5_29_1} Nuclear excitation probability $P_{\mathrm{exc}}(t)$ during a laser pulse for the first electron (left column), the second electron (middle column), and the third electron (right column). Two different laser wavelengths are used, namely, 800 nm (top row) and 400 nm (bottom row). For both wavelengths, the laser pulse has a sine-squared shape with peak intensity 10$^{14}$ W/cm$^2$. Separated contributions from the $M1$ and the $E2$ channels are also shown, as labeled.}
\centering
\end{figure*}

The temporal evolution of the electronic state $\ket{\varphi_{i}(t)}$ obeys the time-dependent ZORA equation
\begin{equation}\label{2022_5_28_1}
  i\frac{\partial }{\partial t}\ket{\varphi_{i}(t)}= \left[ H_\mathrm{ZORA} + H_\mathrm{el}(t) \right] \ket{\varphi_{i}(t)}~.
\end{equation}
The laser electric field is assumed to be linearly polarized along the $z$ axis with amplitude $F_0$, envelope function $f(t)$, and angular frequency $\omega$. Thus, the dipole coupling between the laser and the active electron is written as $H_\mathrm{el}(t) = zF_0f(t)\sin \omega t$.  Equation \eqref{2022_5_28_1} is numerically solved using a generalized pseudospectral method \cite{Yao_1993_CPL, Tong_1997_CP, Chu_2004_PR}. The time propagation of the ZORA equation can be realized using a split-operator
method
\begin{equation}\label{2023_3_3_7}
\begin{split}
   \ket{\varphi_{i}(t+dt)} = &\ \mathrm{exp}({-iH_\mathrm{ZORA}dt/2})\times  \\
    &\ \mathrm{exp}\big[{-iH_\mathrm{el}(t+dt/2)dt}\big]\times \\
    &\ \mathrm{exp}({-iH_\mathrm{ZORA}dt/2})\ket{\varphi_{i}(t)}+O(dt^3)
\end{split}
\end{equation}
where $dt$ is the time step. From the above equation, the time evolution of the wave function from $t$ to $t+dt$ is completed by three steps: (i) evolution for a half-time step $dt/2$ in the energy space spanned by $H_\mathrm{ZORA}$; (ii) evolution for one time step $dt$ in the coordinate space under the influence of the electron-laser coupling $H_\mathrm{el}$; (iii) evolution for another half-time step $dt/2$ in the energy space spanned by $H_\mathrm{ZORA}$. In contrast to Refs. \cite{Yao_1993_CPL, Tong_1997_CP, Chu_2004_PR}, here the wave function at time $t$ is expanded in spherical spinors rather than spherical harmonics for adapting the ZORA Hamiltonian
\begin{equation}\label{2022_3_30_1}
  \ket{\varphi_i(t)}=\sum_{|\kappa|\leq K_{\mathrm{max}}}\ket{R_{\kappa}(r,t)}\ket{\Omega_{\kappa m}(\theta,\phi)},
\end{equation}
where $\ket{R_{\kappa}(r,t)}$ is the (time-dependent) radial wave function, $\ket{\Omega_{\kappa m}(\theta,\phi)}$ is spherical spinors \cite{W.J_Aphy_2007} with quantum number $\kappa$ and magnetic quantum number $m$ , and $K_{\mathrm{max}}$ is an integer to truncate the orbital angular momentum. Here, $m$ is fixed due to $\Delta m=0$ in the linearly polarized laser field. Spherical spinors $\ket{\Omega_{\kappa m}(\theta,\phi)}$ are orthonormal
\begin{equation}\label{2023_3_26_3}
  \braket{\Omega_{\kappa m}}{\Omega_{\kappa^\prime m^\prime}}=\delta_{\kappa \kappa^\prime}\delta_{mm^\prime}
\end{equation}
and they satisfy the eigenvalue equation
\begin{equation}\label{2023_3_26_4}
  (-1-\bm \sigma\cdot\bm L)\ket{\Omega_{\kappa m}(\theta,\phi)}=\kappa\ket{\Omega_{\kappa m}(\theta,\phi)}.
\end{equation}

The calculation is performed in a spherical box with radius 150 a.u. and 300 spatial grid points (nonuniform grid, denser near the origin). The time step is $dt=0.1$ a.u. The orbital angular momentum is truncated at $K_{\mathrm{max}}=80$ in the partial-wave expansion of Eq. \eqref{2022_3_30_1}. A boundary absorbing function $1/[1+\mathrm{exp}(br-r_0)]$ with $b=1.25$ and $r_0=120$ a.u. is used to avoid boundary reflection. The convergency has been carefully checked by varying the calculation parameters. The initial state $\ket{\varphi_i}$ for the first, second, third, and fourth electrons sequentially pulled out by the laser field is the $7s_{1/2}$ orbital of the Th atom, the $7s_{1/2}$ orbital of the Th$^{+}$ ion, the $6d_{5/2}$ orbital of the Th$^{2+}$ ion, and the $5f_{5/2}$ orbital of the Th$^{3+}$ ion, respectively. The magnetic quantum number $m$ is fixed at $1/2$.

\section{Numerical results}

With the time evolution of the electronic states numerically solved, we can calculate the nuclear excitation probability $P_{\mathrm{exc}}(t)$ using Eqs. (\ref{2022_5_24_3}) and (\ref{2022_5_23_1}).
In this section we present $P_{\mathrm{exc}}(t)$ with different laser wavelengths, laser pulse durations, and laser intensities. Comparisons between OE and LDEE channels are also presented.

\subsection{Excitation probability under two wavelengths}

Figure \ref{fig:2022_5_29_1} shows the nuclear excitation probability $P_{\mathrm{exc}}(t)$ during a laser pulse for two different laser wavelengths, namely, 800 nm and 400 nm. For both cases, the laser pulse has a duration of 5 optical cycles with a temporal envelop function $f(t) = \sin^2 (\pi t/NT)$, where $T=2\pi/\omega$ is the period and $N=5$ is the number of optical cycles. The peak intensity of the laser pulse is $10^{14}$ $\mathrm{W}/\mathrm{cm}^2$. This intensity has the ability to drive the outermost three electrons of the Th atom, with negligible effects on the fourth electron, which lies too deeply (A higher intensity is needed to drive this electron, as shown in an example later). Based on the SAE approximation, we calculate separately the first electron, the second electron, and the third electron. For each case, the total excitation probability as well as separated contributions from the $M1$ or $E2$ channels are presented.

With these laser parameters, contributions from the OE channel are found to be 3 to 4 orders of magnitude lower than those from the LDEE channel. This is due to the fact that both 800 nm (photon energy 1.55 eV) and 400 nm (photon energy 3.10 eV) are far off resonance to the nuclear energy gap of 8.28 eV, albeit with a high intensity. Therefore the excitation probabilities presented in Fig. \ref{fig:2022_5_29_1} are almost solely from LDEE. For 800 nm, the end-of-pulse nuclear excitation probability is about $8.2\times10^{-12}$, $8.3\times10^{-13}$, and $6.7\times10^{-14}$ for the first electron, the second electron, and the third electron, respectively. The total excitation probability is about $9.1\times10^{-12}$. For 400 nm, the end-of-pulse nuclear excitation probability is about $1.5\times10^{-11}$, $8.4\times10^{-12}$, and $3.1 \times10^{-12}$ for the first electron, the second electron, and the third electron, respectively. The total excitation probability is about $2.7\times10^{-11}$. This is about three times higher than the 800-nm case.

The relative importance between $M1$ and $E2$ varies from case to case. One sees from the first electron that $E2$ is more important than $M1$ almost for the entire pulse, for both 800 nm and 400 nm. However, the situation reverses for the second electron, where $M1$ dominates during the entire pulse. The situation for the third electron is different for 800 nm and 400 nm. For 800 nm, $M1$ dominates most of the time during the pulse, but near the end of the pulse, the two have almost equal contributions to nuclear excitation. For 400 nm, in contrast, $E2$ dominates for most of the pulse duration except for the initial stage. These results have no simple interpretations but they can be attributed to the dependency of the matrix element of $T_{lm}^\tau$ on the time-dependent electronic states.

\begin{figure}[t!]
\centering
\subfigure{
\centering
\includegraphics[width=8cm]{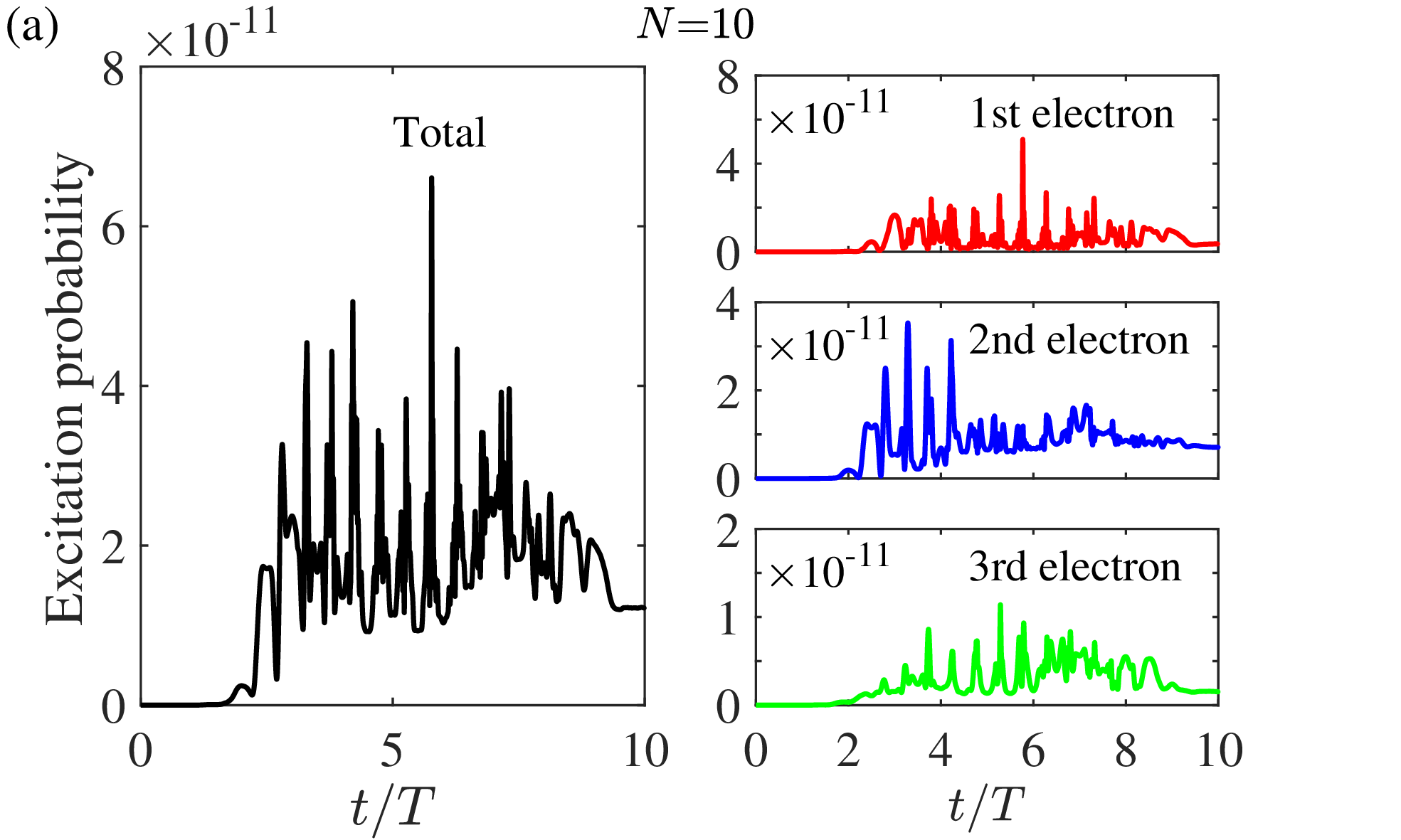}
%\caption{fig1}
}%

\subfigure{
\centering
\includegraphics[width=8cm]{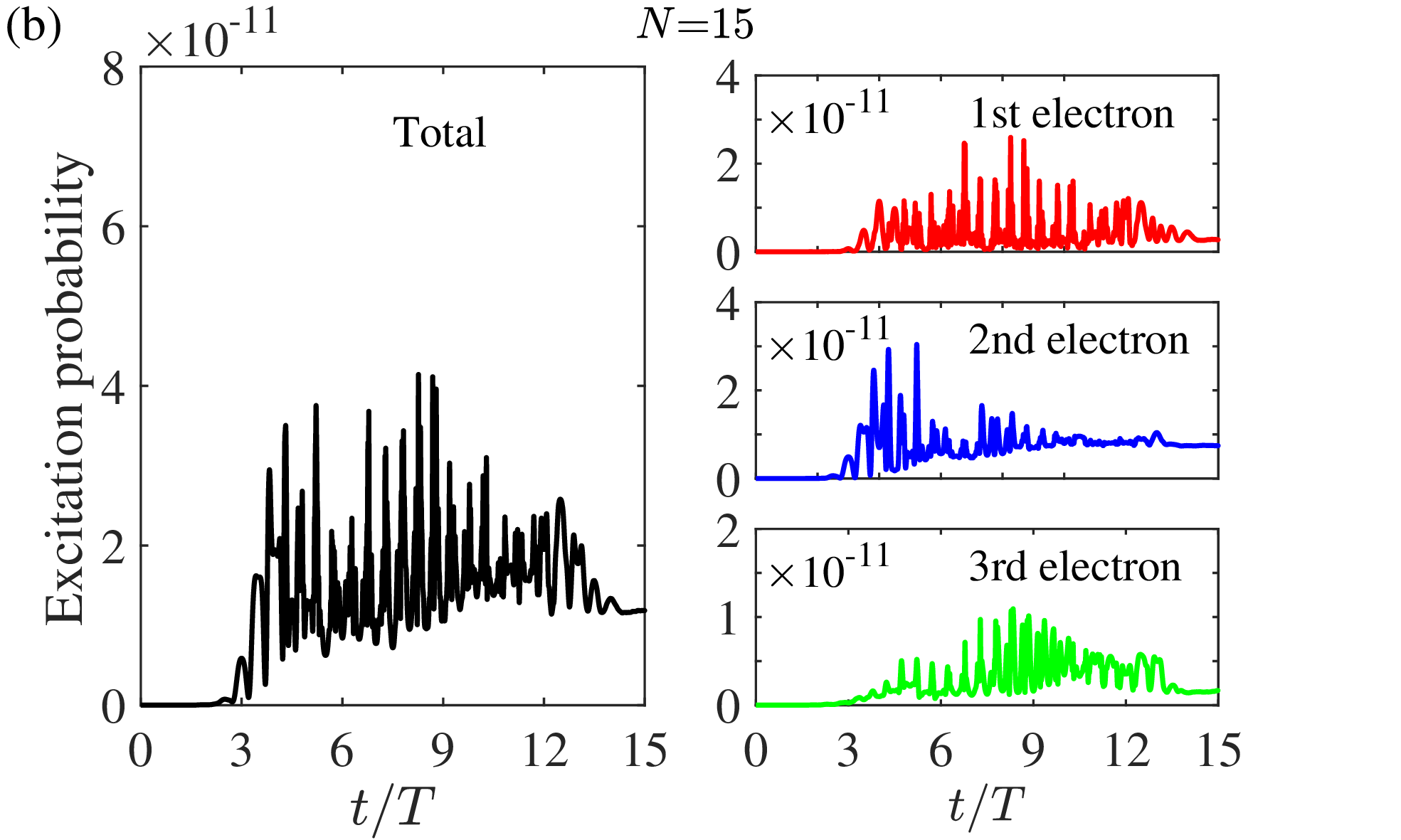}
%\caption{fig2}
}%
\caption{ \label{fig:2023_5_11_1} Nuclear excitation probability $P_{\mathrm{exc}}(t)$ during a laser pulse of duration (a) $N=10$ optical cycles, and (b) $N=15$ optical cycles. The laser wavelength is 400 nm and the peak intensity is $10^{14}$ W/cm$^2$. For each case, separated contributions from the outermost three electrons are also shown.}
\centering
\end{figure}

\subsection{Excitation with different pulse durations}

In Fig. \ref{fig:2023_5_11_1}, we show the nuclear excitation probability $P_{\mathrm{exc}}(t)$ with laser pulses of different durations ($N=10$ and 15 optical cycles). The wavelength and intensity of the laser pulses are fixed at 400 nm and $10^{14}$ W/cm$^2$, so this figure is to be compared with the lower row of Fig. \ref{fig:2022_5_29_1} which is for a shorter pulse of $N=5$. The outermost three electrons contribute to the nuclear excitation and they are calculated separately. 

For the case of $N=10$, the end-of-pulse nuclear excitation probability is about $3.5 \times 10^{-12}$, $7.0\times 10^{-12}$, and $1.5\times10^{-12}$ for the first electron, the second electron, and the third electron, respectively. The total excitation probability is about $1.2\times10^{-11}$.

For the case of $N=15$, the end-of-pulse nuclear excitation probability is about $2.7 \times 10^{-12}$, $7.4\times 10^{-12}$, and $1.7\times10^{-12}$ for the first electron, the second electron, and the third electron, respectively. The total excitation probability is about $1.2\times 10^{-11}$, almost identical to the $N=10$ case. This value is about two times lower than the $N=5$ case shown above (Fig. \ref{fig:2022_5_29_1}).

\begin{figure}[t!]
\centering
\subfigure{
\centering
\includegraphics[width=8cm]{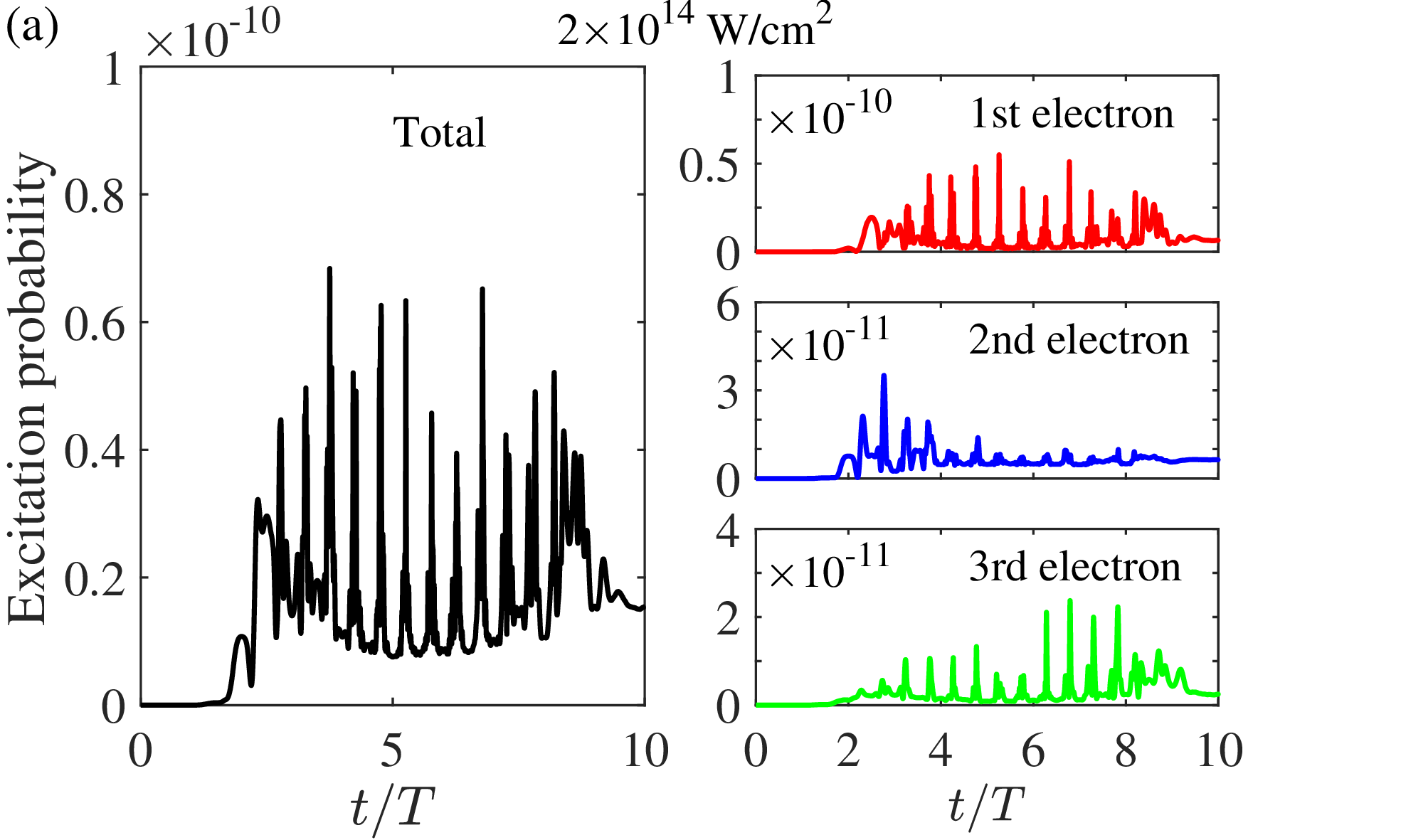}
%\caption{fig1}
}%

\subfigure{
\centering
\includegraphics[width=8cm]{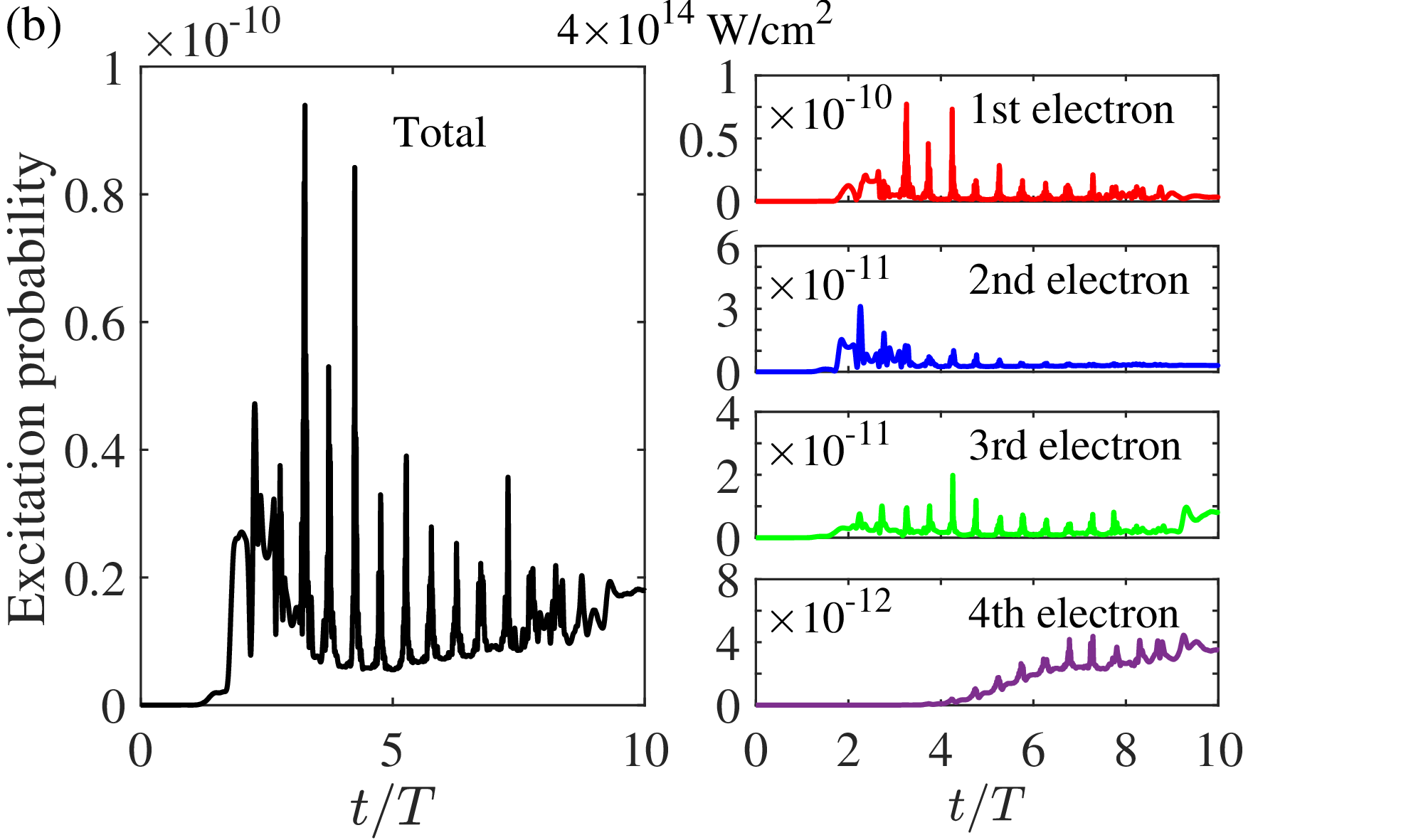}
%\caption{fig2}
}%
\caption{ \label{fig:2023_5_11_2} Nuclear excitation probability $P_{\mathrm{exc}}(t)$ during a laser pulse with intensity (a) $2\times 10^{14}$ W/cm$^2$, and (b) $4\times 10^{14}$ W/cm$^2$. The laser wavelength is 400 nm and the duration is 10 optical cycles. For the higher intensity, the fourth electron starts to contribute to nuclear excitation.}
\centering
\end{figure}

Another noticeable difference is that for $N=5$, the first electron contributes the most to the nuclear excitation, whereas for $N=10$ or 15 the second electron contributes the most. The major difference from the longer pulses is to reduce the contribution from the first electron (from $1.5\times10^{-11}$ for $N=5$, to $3.5\times10^{-12}$ for $N=10$ and $2.7\times10^{-12}$ for $N=15$). Without presenting analyses involving too many details, this pulse-duration effect can be briefly understood as follows: The shorter pulse provides a larger bandwidth such that it drives the first electron to favorable states for the nuclear excitation. Longer pulses reduce the bandwidth and diminish electronic transitions to these favorable states. Nevertheless, the pulse-duration effect is not severe: Under the same peak intensity, pulses with different durations lead to nuclear excitation probabilities within a factor of two or three.

\subsection{Excitation with different laser intensities}

Figure \ref{fig:2023_5_11_2} shows the nuclear excitation probability during two pulses of different peak intensities, namely, $2\times10^{14}$ and $4\times10^{14}$ W/cm$^2$. Both laser pulses have wavelength 400 nm and duration 10 optical cycles.
With intensity $2\times10^{14}$ W/cm$^2$, the first three electrons contribute to the nuclear excitation, whereas with the higher intensity $4\times10^{14}$ W/cm$^2$, the fourth electron starts to contribute to the nuclear excitation.

For the lower intensity, the end-of-pulse nuclear excitation probability is about $6.5\times10^{-12}$, $6.4\times10^{-12}$, and $2.5\times10^{-12}$ for the first electron, the second electron,
and the third electron, respectively. The total nuclear excitation probability is about $1.5\times10^{-11}$.

For the higher intensity, the end-of-pulse nuclear excitation probability is about $3.5\times10^{-12}$, $3.0\times10^{-12}$, $8.0\times10^{-12}$, and $3.5\times10^{-12}$ for the first electron, the second electron, the third electron, and the fourth electron, respectively. The total nuclear excitation probability is about $1.8\times10^{-11}$. From Fig. \ref{fig:2023_5_11_1}(a) and Fig. \ref{fig:2023_5_11_2}, one can see that the total nuclear excitation probability increases with the laser intensity.

\begin{figure}[t!]
\centering
\subfigure{
\centering
\includegraphics[width=5.5cm]{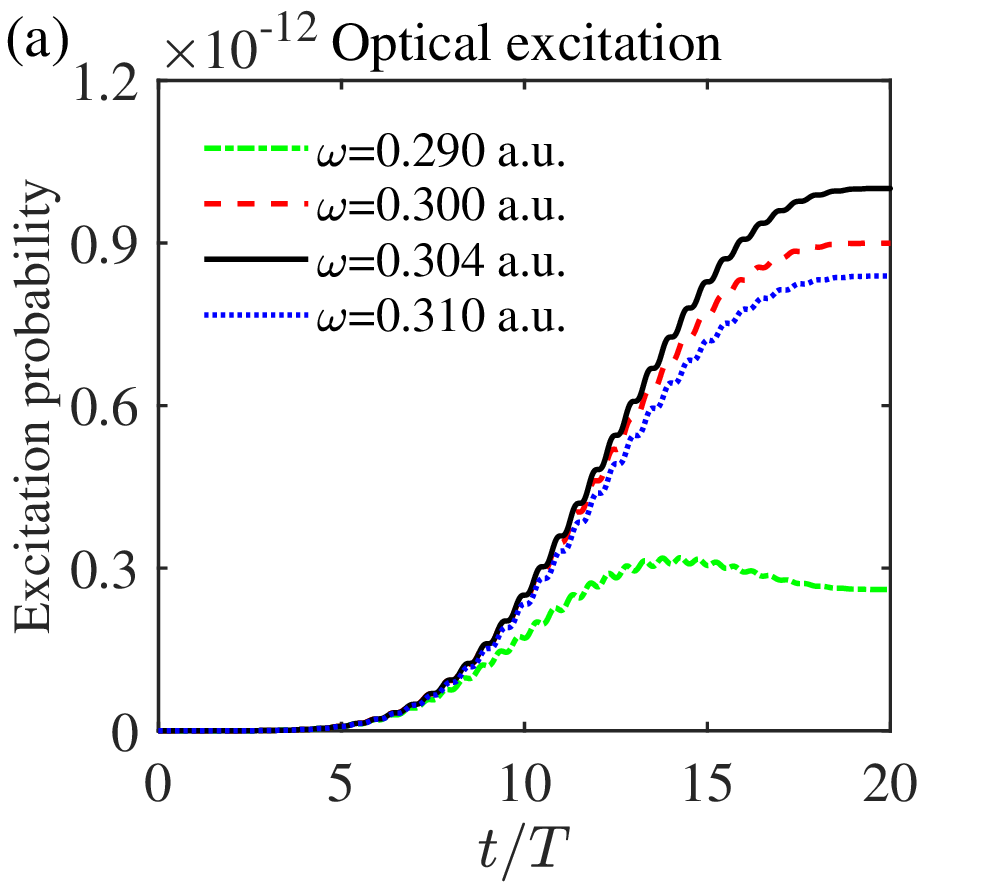}
%\caption{fig1}
}%

\centering
\subfigure{
\centering
\includegraphics[width=5.5cm]{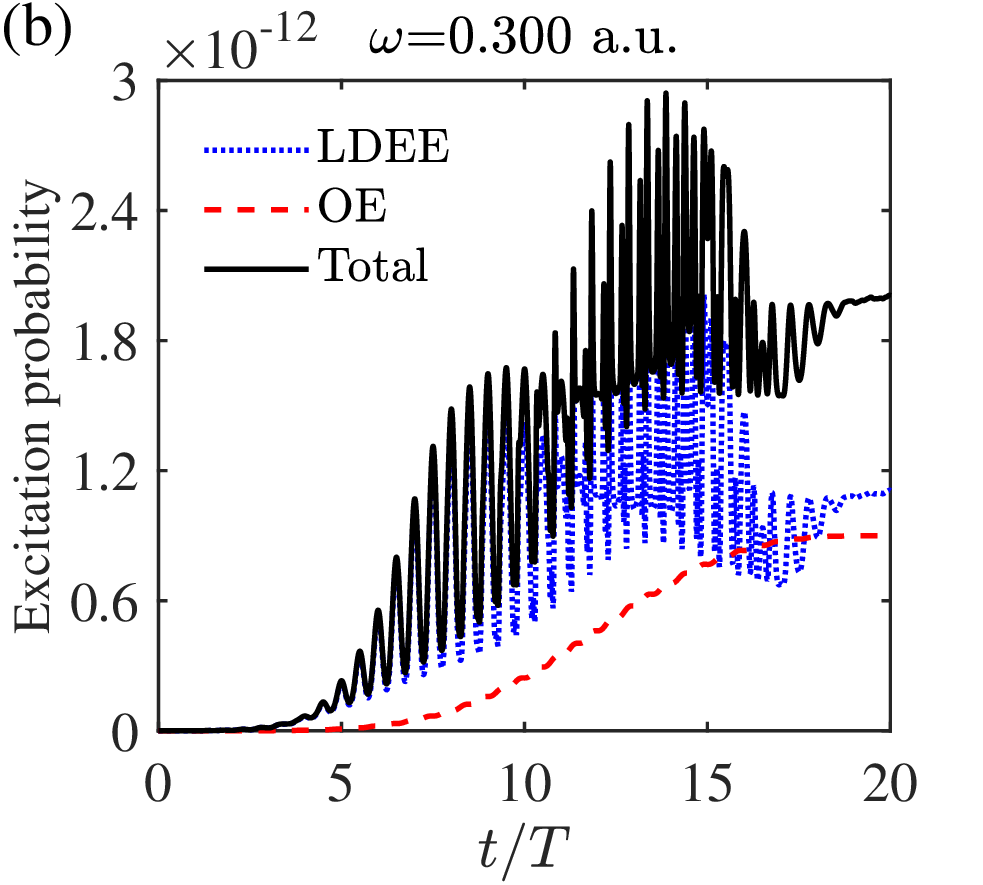}
%\caption{fig1}
}%
\caption{ (a)  Nuclear excitation probability from the OE channel for several laser frequencies around the isomeric resonance (8.28 eV = 0.304 a.u.). The laser peak intensity is 10$^{14}$ W/cm$^2$. (b) Nuclear excitation probability for the Th$^{2+}$ ion with laser frequency 0.300 a.u. (8.16 eV in photon energy) and peak intensity 10$^{14}$ W/cm$^2$. The OE channel and the LDEE channel lead to comparable nuclear excitation probabilities. }
    \label{fig:2022_5_29_2}
\end{figure}

\subsection{OE vs. LDEE channels}

In all the above results, the LDEE channel dominates the nuclear excitation, and the OE channel is weaker by 3 or 4 orders of magnitude. As explained, this is because both 800 nm and 400 nm are far off resonance with the nuclear isomeric energy. The OE channel is important when the laser frequency is close to the isomeric resonance (8.28 eV = 0.304 a.u.). Fig. \ref{fig:2022_5_29_2}(a) shows a few near-resonant examples for peak intensity 10$^{14}$ W/cm$^2$ and pulse duration 20 optical cycles. The nuclear excitation probability can reach about 10$^{-12}$ for exact resonance, but drops in the presence of a detuning.

However, this does not mean that the LDEE channel is negligible. Fig. \ref{fig:2022_5_29_2}(b) shows the comparison between OE and LDEE channels for $\omega=0.300$ a.u., which is very close to the resonant frequency. The calculation is performed with the Th$^{2+}$ ion. One can see that the two channels are comparable in this example, and the LDEE channel is even higher. 
Although there is no easy way of obtaining intense laser pulses with photon energies around 8.28 eV, we want to use this example to emphasize that both OE and LDEE processes exist when the $^{229}$Th atom is radiated by a laser field, and that it would be dangerous to neglect one of them without doing the calculations. This is why a theory including both channels in a single framework is important.

\section{Discussion}

(a) In Ref. \cite{Wang-22}, a semiclassical method is used to calculate the nuclear excitation probability based on the recollision picture (i.e. the RINE process). A probability of $4\times10^{-12}$ is obtained for laser wavelength 800 nm and peak intensity 10$^{14}$ W/cm$^2$. In the quantum calculation given in this paper (Fig. \ref{fig:2022_5_29_1}), an excitation probability of $9.1\times10^{-12}$ is obtained for the same wavelength and peak intensity, although with a shorter pulse duration due to the demanding computational load. On the one hand, this indicates that the semiclassical calculation is fairly accurate, giving at least the right order of magnitude. On the other hand, we emphasize that the quantum calculation includes processes beyond RINE. RINE only includes laser-driven free-free electronic transitions, whereas the quantum calculation also includes free-bound and bound-bound electronic transitions.  

(b) From the quantum calculations, a strong femtosecond laser pulse leads to typical nuclear excitation probabilities on the order of $10^{-11}$ per nucleus. This is to be compared with the 29-keV indirect OE method, which gives an excitation probability on the order of $10^{-11}$ per nucleus per second \cite{Masuda_2019_nature_xray}. That is, a femtosecond laser pulse generates a similar nuclear excitation probability as the (continuous-wave) 29-keV synchrotron radiation does for one second.

(c) Besides the efficiency, our method has the following advantages: (i) Precise knowledge of the isomeric energy is not needed, because the electronic transitions have broad energy distributions which cover the isomeric energy. (ii) Our method is relatively easy to implement experimentally, requiring only tabletop laser systems instead of large facilities. (iii) The excitation is well timed and only happens within the short laser pulse. This may be important for future coherent operations of the excitation process.

(d) We emphasize that laser excitation of $^{229}$Th involves tripartite interactions between the nucleus, the atomic electrons, and the laser field. Although in the current paper we focus on the interaction of $^{229}$Th atoms (ions) with strong femtosecond laser pulses, our theory has more general applicabilities: It provides a general theoretical framework for laser excitation of atomic nuclei.

\section{Conclusion}

In this paper we consider using strong femtosecond laser pulses to excite the $^{229}$Th nucleus. A general quantum mechanical framework is developed to describe the tripartite interaction between the nucleus, the atomic electrons, and the laser field. The nucleus can be excited both by the laser field and by laser-driven electronic transitions. Calculations show that strong femtosecond laser pulses are very efficient in exciting the $^{229}$Th nucleus, leading to excitation probabilities on the order of $10^{-11}$ per nucleus per femtosecond laser pulse. Laser-driven electronic transitions are shown to be more efficient in exciting the nucleus than the laser field itself. The natural and interesting combination between strong-field atomic physics and $^{229}$Th nuclear physics leads to a very efficient nuclear excitation method.

\begin{acknowledgments}
{\it Acknowledgments:} Wu Wang acknowledges discussions with Mrs. Tao Li and Ziqi Liu. This work was supported by NSFC No. 12088101.
\end{acknowledgments}


\begin{thebibliography}{99}

\bibitem{Walker_1999_Nature} P. Walker and G. Dracoulis, Nature \textbf{399}, 35 (1999).
\bibitem{Carroll_2001_Xray_relaaseenergy} J.~J. Carroll, S.~A. Karamian, L.~A. Rivlin, and A.~A. Zadernovsky, Hyperfine Interact. \textbf{135}, 3 (2001).

\bibitem{Zadernovsky2002} A.~A. Zadernovsky and J.~J. Carroll, Hyperfine Interact. \textbf{143}, 153 (2002).
\bibitem{Palffy_2007_PRL_TriggerNEEC} A. P\'alffy, J. Evers, and C.~H. Keitel, Phys. Rev. Lett. \textbf{99}, 172502 (2007).
\bibitem{Yuanbin_PRL_2019} Y. Wu, C.~H. Keitel, and A. P\'alffy, Phys. Rev. Lett. \textbf{122}, 212501 (2019).
\bibitem{Cutler2013} C.~S. Cutler, H.~M. Hennkens, N. Sisay, S. Huclier-Markai, and S.~S. Jurisson, Chem. Rev. \textbf{113}, 858 (2013).

\bibitem{Walker_2020} P. Walker and Z. Podoly{\'{a}}k, Phys. Scr. \textbf{95}, 044004 (2020).

\bibitem{K_R_NPA_1976} L.~A. Kroger and C.~W. Reich, Nucl. Phys. A \textbf{259}, 29 (1976).
\bibitem{Reich-90} C. W. Reich and R. G. Helmer, Phys. Rev. Lett. 64, 271 (1990).
\bibitem{Helmer_PRC_1994} R.~G. Helmer and C.~W. Reich, Phys. Rev. C \textbf{49}, 1845 (1994).
\bibitem{Beck_PRL_2007} B. R. Beck, J. A. Becker, P. Beiersdorfer, G. V. Brown, K. J. Moody, J. B. Wilhelmy, F. S. Porter, C. A. Kilbourne, and
R. L. Kelley, Phys. Rev. Lett. \textbf{98}, 142501 (2007).

\bibitem{Seiferle2019} B. Seiferle et al., Nature \textbf{573}, 243 (2019).
{
\bibitem{Wense2016} L. von der Wense et al., Nature \textbf{533}, 47 (2016).
\bibitem{Thielking2018} J. Thielking, M. V. Okhapkin, P. Glowacki, D. M. Meier, L. von der Wense, B. Seiferle, C. E. D\"ullmann, P. G. Thirolf, and E. Peik, Nature \textbf{556}, 321 (2018).
\bibitem{Minkov_2019_PRL} N. Minkov and A. P\'alffy, Phys. Rev. Lett. \textbf{122}, 162502 (2019).
\bibitem{Yamaguchi_2019_PRL} A. Yamaguchi et al., Phys. Rev. Lett. \textbf{123}, 222501 (2019).
\bibitem{Sikorsky_2020_PRL} T. Sikorsky et al., Phys. Rev. Lett. \textbf{125}, 142503 (2020).
}

\bibitem{Peik_2003} E.~Peik and C. Tamm, Europhys. Lett. \textbf{61}, 181 (2003).
\bibitem{Peik-09} E. Peik, K. Zimmermann, M. Okhapkin, and C. Tamm, in {\it Proc. 7th Symp. on Frequency Standards and Metrology}, pages 532-538 (edited by L. Maleki, World Scientific, 2009).
\bibitem{Rellergert-10} W. G. Rellergert, D. DeMille, R. R. Greco, M. P. Hehlen, J. R. Torgerson, and E. R. Hudson, Phys. Rev. Lett. \textbf{104}, 200802 (2010).
\bibitem{Campbell_2012_PRL_clock} C.~J. Campbell, A.~G. Radnaev, A. Kuzmich, V.~A. Dzuba,
V.~V. Flambaum, and A. Derevianko, Phys. Rev. Lett. \textbf{108}, 120802 (2012).

\bibitem{Flambaum-06} V. V. Flambaum, Phys. Rev. Lett. \textbf{97}, 092502 (2006).
\bibitem{Berengut-09} J. C. Berengut, V. A. Dzuba, V. V. Flambaum, and S. G. Porsev, Phys. Rev. Lett. \textbf{102}, 210801 (2009).
\bibitem{Fadeev-20} P. Fadeev, J. C. Berengut, and V. V. Flambaum, Phys. Rev. A \textbf{102}, 052833 (2020).

\bibitem{Jeet_PRL_2015} J.~Jeet, C. Schneider, S. T. Sullivan, W. G. Rellergert, S. Mirzadeh, A. Cassanho, H. P. Jenssen, E. V. Tkalya, and
E. R. Hudson, Phys. Rev. Lett. \textbf{114}, 253001 (2015).

\bibitem{Yamaguchi_NewJPhys_2015} A.~Yamaguchi, M. Kolbe, H. Kaser, T. Reichel, A. Gottwald, and E. Peik, New J. Phys. \textbf{17}, 053053 (2015).
\bibitem{Peik-15} E. Peik and M. Okhapkin, C. R. Phys. \textbf{16}, 516 (2015).
\bibitem{Stellmer_PRA_2018} S. Stellmer, G. Kazakov, M. Schreitl, H. Kaser, M. Kolbe, and T. Schumm, Phys. Rev. A \textbf{97}, 062506 (2018).

\bibitem{Masuda_2019_nature_xray} T.~Masuda et~al., Nature \textbf{573}, 238 (2019).

\bibitem{Tkalya-00} E. V. Tkalya, A. N. Zherikhin, and V. I. Zhudov, Phys. Rev. C \textbf{61}, 064308 (2000).

\bibitem{Tkalya-20} E. V. Tkalya, Phys. Rev. Lett. \textbf{124}, 242501 (2020).
\bibitem{Zhang-22} H. Zhang, W. Wang, and X. Wang, Phys. Rev. C \textbf{106}, 044604 (2022).

\bibitem{Zhang-23} H. Zhang and X. Wang, Front. Phys. 11:1166566 (2023).

\bibitem{Tkalya-92} E. V. Tkalya, JETP Lett. \textbf{55}, 212 (1992).
\bibitem{Porsev_2010_PRL_TEB} S. G. Porsev, V. V. Flambaum, E. Peik, and C. Tamm, Phys. Rev. Lett. \textbf{105}, 182501 (2010).
\bibitem{Borisyuk_2019_PRC_EBcontinuum} P. V. Borisyuk, N. N. Kolachevsky, A. V. Taichenachev, E. V. Tkalya, I. Yu. Tolstikhina, and V. I. Yudin, Phys. Rev. C \textbf{100}, 044306 (2019).
\bibitem{Bilous_2020_PRL_Ions} P.~V.~Bilous, H.~Bekker, J.~C.~Berengut, B.~Seiferle, L.~von der~Wense, P.~G.~Thirolf, T.~Pfeifer, J.~R.~Crespo L\'opezUrrutia, and A.~P\'alffy, Phys. Rev. Lett. \textbf{124}, 192502 (2020).
\bibitem{Nickerson_2020_PRL_Nuclear} B. S. Nickerson, M. Pimon, P. V. Bilous, J. Gugler, K. Beeks, T. Sikorsky, P. Mohn, T. Schumm, and A. P\'alffy,
Phys. Rev. Lett. \textbf{125}, 032501 (2020).

\bibitem{Bilous_2018_PRC} P.~V. Bilous, N. Minkov, and A. P\'alffy, Phys. Rev. C \textbf{97}, 044320 (2018).

\bibitem{Feng-22} J. Feng et al. Phys. Rev. Lett. \textbf{128}, 052501 (2022).

\bibitem{Qi-23} J. Qi, H. Zhang, and X. Wang, Phys. Rev. Lett. \textbf{130}, 112501 (2023).

\bibitem{Wu_PRL_2021} W. Wang, J. Zhou, B. Liu, and X. Wang, Phys. Rev. Lett. \textbf{127}, 052501 (2021).

\bibitem{Wang-22} X. Wang, Phys. Rev. C \textbf{106}, 024606 (2022).


\bibitem{Kulander_Plenum_1993} K. C. Kulander, K. J. Schafer, and J. L. Krause, in Super-Intense
Laser-Atom Physics, edited by B. Piraux, A. L'Huillier, and K.
Rzazewski (Plenum, New York, 1993).
\bibitem{Schafer-93} K. J. Schafer, B. Yang, L. F. DiMauro, and K. C. Kulander, Phys. Rev. Lett. \textbf{70}, 1599 (1993).
\bibitem{Corkum-93} P. B. Corkum, Phys. Rev. Lett. \textbf{71}, 1994 (1993).


\bibitem{McPherson_1987} A. McPherson, G. Gibson, H. Jara, U. Johann, T. S. Luk, I.
A. McIntyre, K. Boyer, and C. K. Rhodes, J. Opt. Soc. Am. B \textbf{4}, 595 (1987).
\bibitem{Ferray_1988} M. Ferray, A. L'Huillier, X. F. Li, L. A. Lompre, G.
Mainfray, and C. Manus, J. Phys. B \textbf{21}, L31 (1988).
\bibitem{Seres2005} J.~Seres,  E. Seres, A. J. Verhoef, G. Tempea, C. Streli, P. Wobrauschek, V. Yakovlev, A. Scrinzi, C. Spielmann, and F. Krausz, Nature \textbf{433}, 596 (2005).


\bibitem{Walker-94} B. Walker, B. Sheehy, L. F. DiMauro, P. Agostini, K. J. Schafer, and K. C. Kulander, Phys. Rev. Lett. \textbf{73}, 1227 (1994).
\bibitem{Palaniyappan-05} S.~Palaniyappan, A. DiChiara, E. Chowdhury, A. Falkowski, G. Ongadi, E. L. Huskins, and B. C. Walker, Phys. Rev. Lett. \textbf{94}, 243003 (2005).
\bibitem{Becker-12} W. Becker, X. Liu, P. J. Ho, and J. H. Eberly, Rev. Mod. Phys. \textbf{84}, 1011 (2012).

\bibitem{Morishita-PRL-2008} T. Morishita, A. T. Le, Z. J. Chen, and C. D. Lin, Phys. Rev. Lett. \textbf{100}, 013903 (2008).
\bibitem{Blaga-Nature-2012} C. I. Blaga et al., Nature (London) \textbf{483}, 194 (2012).
\bibitem{Wolter-Science-2016} B. Wolter et al., Science \textbf{354}, 308 (2016).

\bibitem{Krausz-09} F. Krausz and M. Ivanov, Rev. Mod. Phys. \textbf{81}, 163 (2009).
\bibitem{Zhao-12} K. Zhao, Q. Zhang, M. Chini, Y. Wu, X. Wang, and Z. Chang, Opt. Lett. \textbf{37}, 3891 (2012).
\bibitem{Li-17}  J. Li et al., Nat. Commun. \textbf{8}, 186 (2017).
\bibitem{Gaumnitz-17} T. Gaumnitz, A. Jain, Y. Pertot, M. Huppert, I. Jordan, F. Ardana-Lamas, and H. J. W\"orner, Opt. Express \textbf{25}, 27506 (2017).


\bibitem{Andreev_PRA_2019} A. V. Andreev, A. B. Savel'ev, S. Yu. Stremoukhov, and O. A. Shoutova, Phys. Rev. A \textbf{99}, 013422
(2019).
\bibitem{Wu_JPB_2021} W. Wang, H. Zhang, and X. Wang, J. Phys. B 54, 244001 (2021).


\bibitem{Schwartz_1957} C. Schwartz, Phys. Rev. \textbf{97}, 380 (1955).
\bibitem{W.J_Aphy_2007} W. R. Johnson, \newblock {\it Atomic Structure Theory: Lectures on Atomic Physics} (Springer, 2007).

\bibitem{Minko_PRC_2021} N. Minkov and A. P\'alffy, Phys. Rev. C \textbf{103}, 014313 (2021).

\bibitem{ADK} M. V. Ammosov, N. B. Delone, and V. P. Krainov, Sov. Phys. JETP \textbf{64}, 1191 (1986).
\bibitem{Kulander-87} K. C. Kulander, Phys. Rev. A \textbf{35}, 445(R) (1987).
\bibitem{Awasthi-08} M. Awasthi, Y. V. Vanne, A. Saenz, A. Castro, and P. Decleva, Phys. Rev. A \textbf{77}, 063403 (2008).
\bibitem{Le-16} A.-T. Le, H. Wei, C. Jin, and C. D. Lin, J. Phys. B \textbf{49}, 053001 (2016).

\bibitem{Chang_1986} C.~Chang, M.~Pelissier, and Ph.~Durand, Phys. Scr. \textbf{34}, 394 (1986).
\bibitem{Lenthe_1993} E.~van Lenthe, E.~J. Baerends, and J.~G. Snijders, J. Chem. Phys. \textbf{99}, 4597 (1993).
\bibitem{Lenthe_1994} E.~van Lenthe, E.~J. Baerends, and J.~G. Snijders, J. Chem. Phys. \textbf{101}, 9783 (1994).

\bibitem{radial_2019} F. Salvat and J.~M. Fern\'andez-Varea, Comput. Phys. Commun. \textbf{240}, 165 (2019).

\bibitem{Yao_1993_CPL} G. Yao and S. I. Chu, Chem. Phys. Lett. \textbf{204}, 381 (1993).
\bibitem{Tong_1997_CP} X. M. Tong and S. I. Chu, Chem. Phys. \textbf{217}, 119 (1997).
\bibitem{Chu_2004_PR} S. I. Chu and D. A. Telnov, Phys. Rep. \textbf{390}, 1 (2004).



\end{thebibliography}
\end{document}